\documentclass[%
 reprint,
nofootinbib,
 amsmath,amssymb,
 aps,
]{revtex4-2}
\usepackage{graphicx}% Include figure files
\usepackage{dcolumn}% Align table columns on decimal point
\usepackage{bm}% bold math
\usepackage{xcolor}
\usepackage{hyperref}% add hypertext capabilities
\usepackage{Tomspeak}
\begin{document}

\preprint{APS/123-QED}

\title{Sub-singularities for shaping thin sheets}% Force line breaks with \\
\thanks{A footnote to the article title}%
\author{Thomas A. Witten}%
 \email{t-witten@uchicago.edu}
\author{Anna Movsheva}
\affiliation{Department of Physics and the James Franck Institute, The University of Chicago, Chicago, IL, 60637, USA
}
%---if Efi is an author,.....
%\author{Efi Efrati}
%\affiliation{Physics Department, Weizmann Institute of Science, Rehovot, ISRAEL
%}
%

\date{\today}% It is always \today, today,
%  but any date may be explicitly specified

\begin{abstract}
The principles behind the sharp, singular structures in a crumpled sheet are well understood.  Here we discuss more general ways of exploiting such sharp structures to control the shape of a sheet by deforming or forcing it elsewhere.  Often, the induced shape leads to further sharp structures---``sub-singularities".  Though weaker and softer than the primary  singularities, they nevertheless provide robust ways of shaping a sheet.  In simple cases we understand the reason for these and their strength.  This paper surveys a broad range of other sub-structure phenomena and reports recent progress in understanding some of them.  
\end{abstract}
% \begin{description}
% \item[Usage]
% Secondary publications and information retrieval purposes.
% \item[Structure]
% You may use the \texttt{description} environment to structure your abstract;
% use the optional argument of the \verb+\item+ command to give the category of each item. 
% \end{description}

%Test the effects of betas on the b's obtained

%Find some way to mark the changes in the latex file

%How much does the chi square change in FIG 5 if we do not have an intercept

%\keywords{Suggested keywords}%Use showkeys class option if keyword
                              %display desired
\maketitle

%\tableofcontents

\section{Introduction} \label{sec:Introduction}

In the 1990's soft-matter physicists began applying their scaling approach to reveal novel properties of elastic sheets in the limit of small thickness\cite{pomeau1995papier,Lobkovsky1995Scaling-Propert,Amar:1997fk,Cerda:1998cj}.  More recently, unexpected and general discoveries about thin sheets emerged from the lab at École de Physique et Chimie (ESPCI) led by Beno\^it Roman and José Bico\cite{Py:2007nr, roman2003oscillatory, Siefert:2019uo} under the encouragement of \'Etienne Guyon. 
By now the study of thin sheets as soft matter has grown to encompass many kinds of deformation,\cite{Hutton2024Buckling-mediat,Cerda:1999fr, Davidovitch:2011wq, bayart2010tearing, Aharoni2017The-smectic-ord,  Hure:2011fk, pomeau1997plates, Vandeparre:2011cs, Duncan:1982dw, Meeussen2023Multistable-she, tani2024curvy} many materials\cite{Armon2011Geometry-and-Me,  Vella:2015qy, Meeussen2023Multistable-she},  many kinds of forcing\cite{Sharon:2002kx,Box2020Dynamic-Bucklin, Pal2024Faceted-wrinkli, He:2023rm} and boundary effects \cite{Tobasco:2022xe}.

In this paper we consider a small subset of these phenomena. We consider {\em singular} structures, with strong curvature confined to an indefinitely  small fraction of the sheet.  This contrasts with the smooth structure of a wrinkled surface.  We consider {\em spontaneously formed} structures, in contrast to explicitly folded or cut structures.  Such folds or cuts are only used as a means to induce the spontaneously formed structures of interest.  We confine our treatment to flat, Euclidean sheets without exploring their generalization to shells or other structures with non-flat metrics \cite{Sharon:2007le, Kim2012Designing-Respo}.  

Why should a sheet develop these concentrated deformations whose characteristic sizes go to zero with the thickness of the sheet? 
The answer lies in the competition between two forms of elastic energy in a thin sheet.  On the one hand, energy is required to stretch or shear or compress the sheet, even when it is confined to its undistorted two-dimensional plane.  On the other hand, energy is required to bend the sheet out of this plane.  The energy costs of these two forms of deformation depend differently on the thickness.  Thus the shape resulting from competition between these two, likewise depends on thickness.  In the next section we review the simplest and strongest instance of this competition.  This ``stretching ridge" is present in every crumpled sheet.  

In this paper we focus on other forms of concentrated deformation that are qualitatively weaker than the primary stretching-ridge.  These 
``sub-singularities" concentrate deformation, without limit as thickness goes to zero, but much less than a comparable stretching ridge. 
 Little has been done to establish their scaling properties with thickness.  In some cases, the resulting shape in the zero-thickness limit is not even established.  We begin with a review of how singular structures arise.

\section{Thin-sheet singularities}\label{sec:thinsheet}

\begin{figure}[ht]
	\includegraphics[width = 1\linewidth] {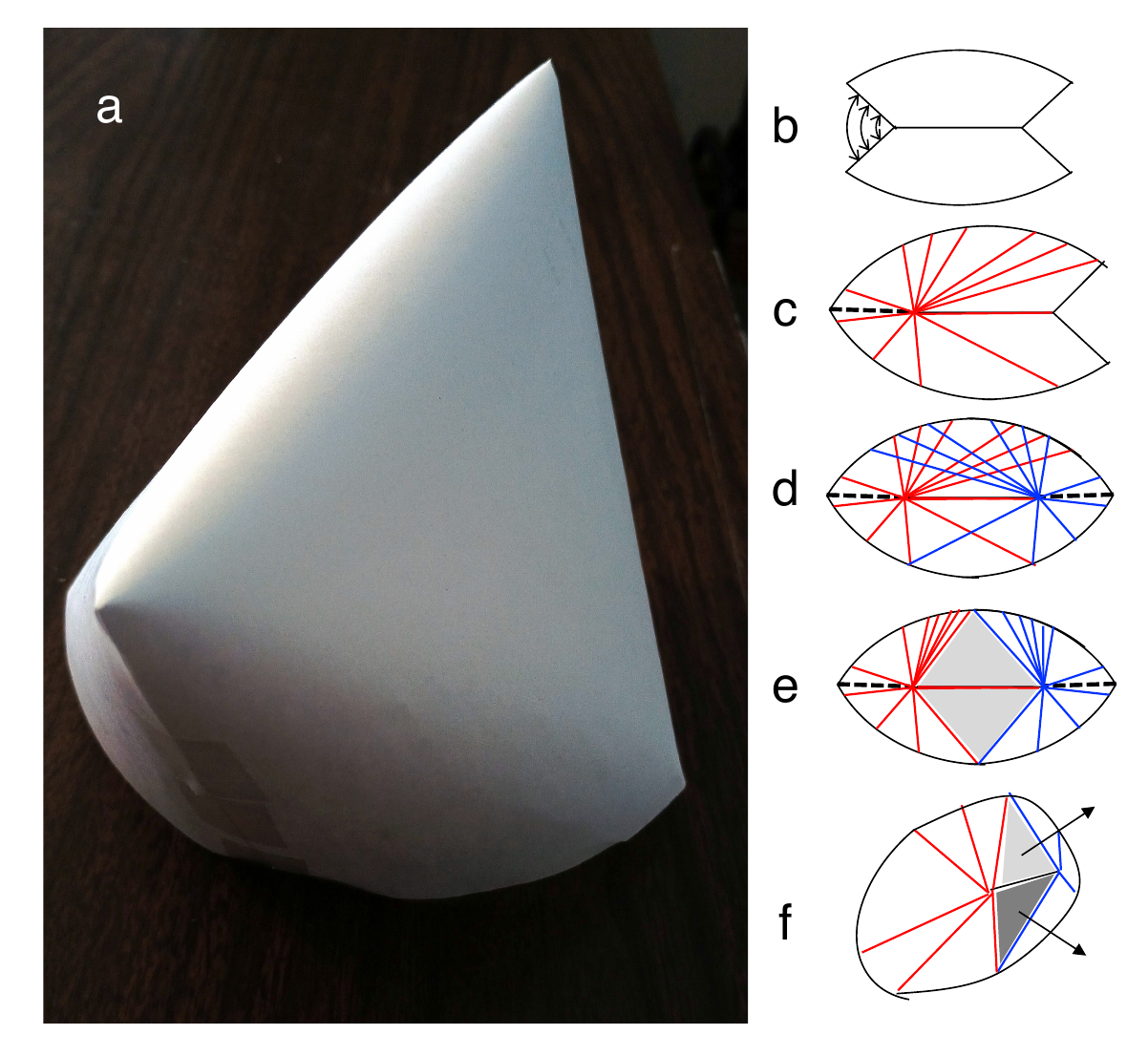}
    \caption {Bag geometry to illustrate stretching ridge mechanism.  a) Photograph of a sheet of office paper with top and bottom halves joined on both left and right sides. b) Forming a cone by joining cut edges, c) Generator lines in the resulting cone.  d) Superimposed generator lines of two cones combined, showing crossing generators, e) The facets (shaded region) resulting from displacing the generators to avoid intersection.  f) Oblique view showing the resulting angle between the facets.}
    \label{fig:bag}
\end{figure}
Figure \ref{fig:bag} is a photo of a piece of paper whose top and bottom halves have been joined at the left and right edges to form a bag shape.  If only one side had been joined, the resulting shape would have been a cone.  But the shape with both edges joined is much different: the curvature is concentrated into a narrow ridge at the top.  

The conical shape (Fig. \ref{fig:bag}c) is understandable via geometry.  A cone is the simplest case of an ``isometric deformation."  If a flat ideal sheet is bent into an ideal cone, the distance along the surface between any two nearby points does not change.  That is, there is no stretching, shear, or compression deformation anywhere.  There is a simple geometric condition that assures this isometry.  This ``developability condition"\cite{Witten:2007wq} restricts the curvature that the surface can have at every point: there must be at least one direction with {\em no} curvature; any curvature must be perpendicular to that direction.  Further, if there is curvature at a point, then the line joining the uncurved directions from the point must be straight.  This ``generator" line from each point must extend as far as the sheet remains isometric.  In a cone, these generators are the radial lines pointing from any point to the cone apex.  The energy cost of forming the cone consists entirely of bending energy and is proportional to the ``bending stiffness" $\kappa$ \cite{[{Denoted $D$ in }] [{, Chap. II.}] Landau:1986sf} of the sheet. 

When both edges are joined (Fig. \ref{fig:bag}d), the conical shape is no longer sustainable.  The original generators from the left apex would intersect those of the right apex.  The result is that generators are excluded from a triangular region between the apexes, forming a flat ``facet."   This faceted and folded structure is clearly singular: the fold has infinite curvature, in this zero-thickness limit.

In a sheet of nonzero thickness, the singularity can be softened by allowing stretching.  If the curved region is smoothed over a finite width $w$, one may readily estimate the resulting stretching and bending energies\cite{Lobkovsky1995Scaling-Propert} to find an optimum $w$ for which the total energy is smallest. Since the two energies have different thickness dependences, the optimal width also varies with thickness $t$. One finds\cite{Lobkovsky1995Scaling-Propert} that $w/X$ varies as   $(t/X)^{1/3}$, where $X$ is the length of the facet boundary.  The optimal energy, in units of $\kappa$,  also increases as $X/w$  \cite{Lobkovsky1995Scaling-Propert}.
Further, these ``stretching ridge" structures dominate the shape and energy of crumpled sheets\cite{Witten:2007wq}.  The sub-singularities we now consider are to be viewed in contrast to these ridges.

\begin{figure}[ht]
\includegraphics[width = 1\linewidth] {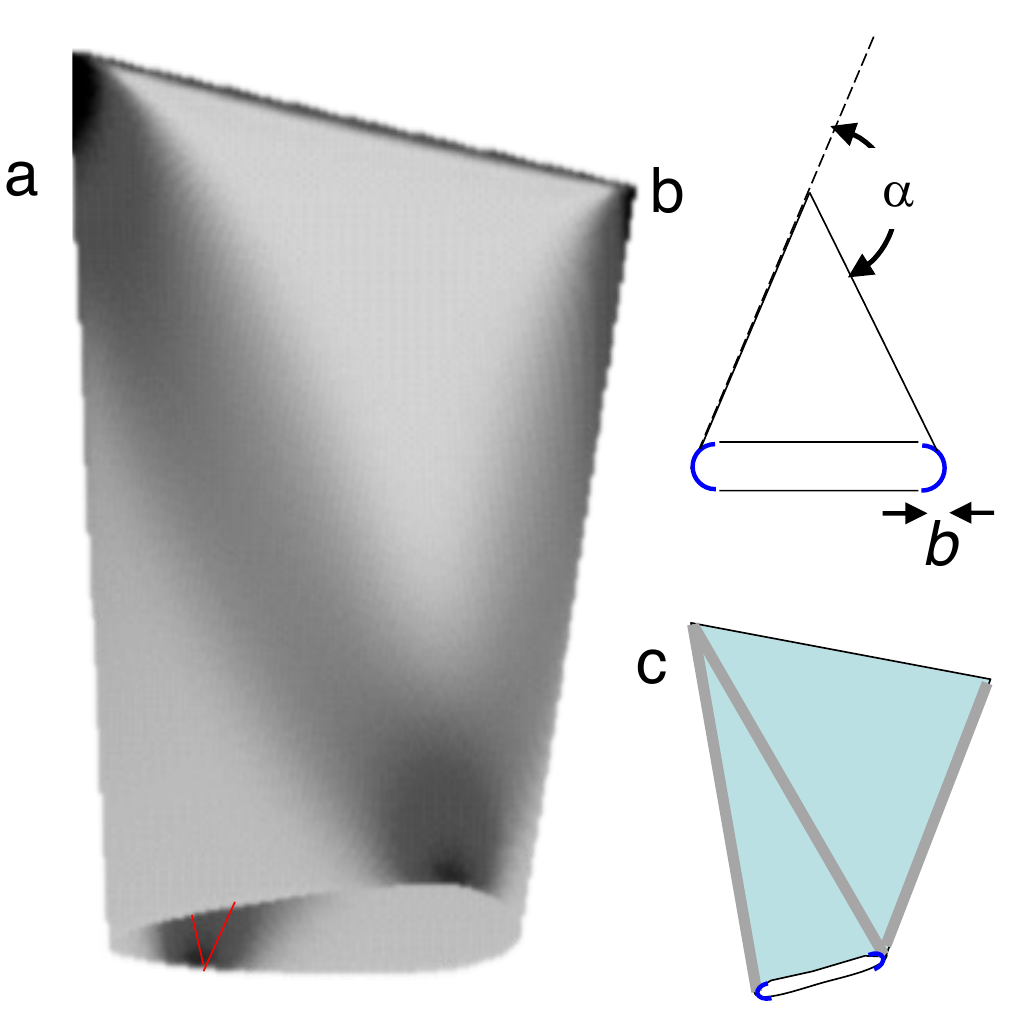}
 \caption{Induced boundary ridge in a bag shaped sheet such as Fig. \ref{fig:bag}a. a) Numerical counterpart of Fig. \ref{fig:bag}a from Ref. \cite{Witten:2009kq, Lobkovsky1996Structure-of-Cr}. Stretching ridge of length $X$ is along the top. Shading is proportional to the elastic energy density. b) Sketch of the anticipated thin-sheet limit in end view along the stretching ridge, showing facet bending angle $\alpha$ near its minimum $\alpha^*$. Bottom shows free boundary with tightly curved ends of radius $b$. c) Sketch of oblique view.  Strongly bent edges from free boundary to cone edges are indicated by gray lines from free boundaries to cone vertices.}
 \label{fig:boundaryridge}
 \end{figure}
\section{induced sub-singularity}\label{sec:induced_bag}

One feature of Fig. \ref{fig:bag} has no obvious explanation via the reasoning above.  The boundaries of the facets show a band of concentrated deformation energy. This concentration has a natural explanation as a side effect of the central stretching ridge, which dominates the total elastic energy.  The ridge energy strongly favors a small bending angle $\alpha$ \cite{Lobkovsky1996Boundary}  between the two facets.  Since this energy becomes arbitrarily strong\footnote{in units of the bending stiffness $\kappa$} in the thin limit, we consider the minimum angle that $\alpha$ can assume, denoted $\alpha^*$.  This minimization requires that the tips of the two facets be as far apart as possible, thus stretching the free boundary into a pair of straight lines perpendicular to the ridge, This turns the entire sheet into a flat-sided tetrahedron, as illustrated in Fig. \ref{fig:boundaryridge}.  If $\alpha$ is slightly more than this $\alpha^*$, the region near the tip is bent into a tight semicircle of some small radius $b$ and curvature of order $1/b$.  The generators for this semicircle end at the two original apexes at the ends of the stretching ridge, forming narrow partial cones.  These induced cones become indefinitely narrow in the thin limit, as $\alpha \goesto \alpha^*$.  

Ref. \cite{Witten:2009kq} calculates how the width $b$ of the induced cones varies with the thickness.  To summarize this argument, we shall express all lengths in units of the ridge length $X$.  We shall also suppose that the aspect ratio of the initial sheet is some fixed constant of order unity. Then $\alpha^*$ is also of order unity.  Now all elastic energies can be expressed in terms of the bending modulus $\kappa$ and the thickness $t$.  

To determine the optimal $\alpha$, we must balance the ridge-energy cost of increasing $\alpha$ beyond $\alpha^*$ against the diverging cost of decreasing the cone widths $b$ to zero.  The ridge energy varies strongly with $\alpha$.  It varies as a power of $\alpha$, specifically the 7/3 power.\cite{Lobkovsky1996Boundary}.
% can be derived as a consequence of the virial theorem, I believe.  We shall assume such a power for our estimate.  
As for the cone energy $E_c$, it is proportional to the bending modulus $\kappa$ and inversely proportional to the radius $b$.  Its conical shape gives a logarithmic divergence to $E_c$. This logarithmic factor (denoted $L$) depends on an inner length no smaller than $t$ within which the cone is no longer developable.  This factor does not depend on $b$.  Thus, 
\begin{equation}
E_c \goesto C ~L~\kappa /b 
,\end{equation}
where $C$ is a numerical factor of order unity.  

As $b$ increases from 0, $\alpha$ necessarily increases from $\alpha^*$.  The length $Y$ of the free boundary must remain unchanged as $b$ increases from zero.  A little geometric reasoning shows that the $\alpha$ increases in proportion to $b$; so that 
\begin{equation}
db/d\alpha = C'
,\end{equation}
where $C'$ is another  constant of order unity.

The competing ridge energy is a power law $E_r(\alpha)\goesas \alpha^p$\cite{Lobkovsky1996Boundary}. Thus, its derivative $dE_r/d\alpha$ can be expressed
\begin{equation}
\left .{dE_r\over d\alpha} \right |_{\alpha^*} = p {E_r{}^*\over \alpha^*}
\end{equation}
The optimal $\alpha$ is that for which $d(E_c + E_r)/d\alpha = 0$.  We can express $d E_c/d\alpha$ as 
\begin{equation}
{d E_c\over d\alpha} = {d E_c\over db}~ {db\over d\alpha} = C' {d E_c\over db} = -C' ~C~L~ \kappa /b^2
.\end{equation}
The optimal $b$ thus satisfies
\begin{equation}
C' ~C~L~ \kappa /b^2 = p E_r{}^*/\alpha^*
\label{eq:balance}
\end{equation}

We may directly compare the sharpness of these induced cones with that of the dominant ridge singularity. As noted above, $E_r = C'' \kappa/w$ for some constant $C''$.  Thus, Eq. \ref{eq:balance} becomes
\begin{equation}
C' ~C~ \kappa /b^2 = C'' ~p~ \kappa /(w ~\alpha^*)
\end{equation}
Thus $b \proportionalto \sqrt w$. Since $w\goesto 0$ with thickness, so must $b$.  However $b$ is qualitatively larger and less sharp than the ridge.  It is a subordinate singularity.  

Intriguingly, one finds similar cases of mild singularities induced by various types of deformation in thin sheets.  We now review some of these.

\section{Menagerie}\label{sec:Menagerie}
\begin{figure}[ht]
	\includegraphics[width = 1\linewidth] {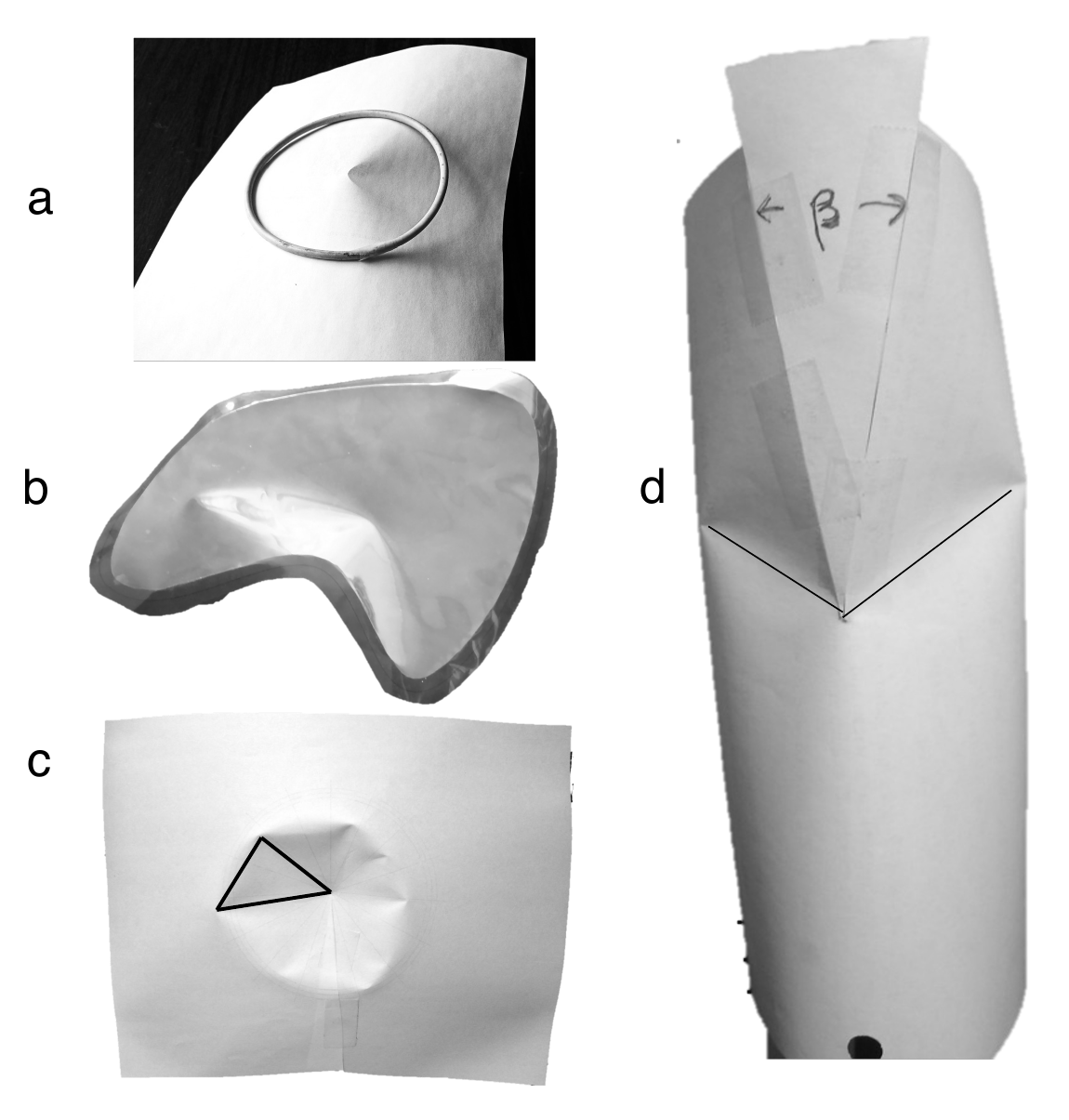}
    \caption {Examples of sub-singularities. a) D-cone in paper by forcing a ring down onto a pointed object. b) D-cones in a plastic sheet deformed by a curved crease near the boundary.  c) Cone in a paper sheet whose vertex region was forced to invert. Lines indicate a facet boundary from the cone vertex. d) Paper cylinder with a wedge of angle $\beta$ inserted at the top, as described in text\footnote{
    The authors thank Efraim Efrati for identifying this version of the sub-singularity.}.  }
    \label{fig:menagerie}
\end{figure}
Deforming thin sheets can produce focused structures that seem as intrinsic as the stretching ridges of Sec. \ref{sec:thinsheet}, but are ill-understood.  In this section we survey some of these, to show their variety and generality.  

The first of these, shown in Fig. \ref{fig:menagerie}a, is made by forcing a flat sheet into a ring \cite{Cerda:1999fr}.  As thickness goes to zero, the overall shape converges to a conical structure, with generators pointing to a common center. This outer region thus approaches a developable surface, whose shape is independent of thickness and is described entirely by bending deformation.  This developable ``d-cone" shape cannot persist to the common center; one expects an inner structure in which stretching occurs.  However, energy-optimization arguments like those of Sections \ref{sec:thinsheet} and \ref{sec:induced_bag} do not seem adequate to explain it.\cite{Witten:2007wq}.  Empirically, the inner structure for a ring of size $X$ and thickness $t$ has a size $h$ varying as $(t/X)^{1/3}$, like the width $w$ of a stretching ridge.  But there is no apparent singular locus for such a ridge.  

Such d-cone shapes are pervasive.  Fig. \ref{fig:menagerie}b shows two made by creating a crease\cite{Cerda:1999fr} in the shape of a closed loop around the boundary of a sheet. The figure shows two d-cones in the interior, controlled by the enclosing crease.

Fig. \ref{fig:menagerie}c shows an ordinary cone in a sheet of paper.  The center region has been inverted by forcing the tip into the cone.  Instead of the anticipated circular crease separating the inner and outer regions, the crease is broken into segments separated by sharp, d-cone-like crescents. The inner region has changed from its original circular shape into a faceted surface.  This kind of faceting must be pervasive in many variants.  The puzzle is how the sheet determines the number of facets according to its thickness.  

Fig. \ref{fig:menagerie}d shows another form of emergent d-cone.  It arises in cases where a tube must be enlarged to match onto a larger tube. It is made by making an axial cut into  the smaller cylindrical tube and then inserting a triangular wedge to between the two cut edges, thus forcing them apart.  As the figure shows, the inserted wedge has caused the upper part to splay out and has induced a d-cone-like singularity on either side.  The d-cones and the grooves leading from each d-cone to the wedge tip appear to be a necessary consequence of the mismatch between the wedged sheet on the right and the cylinder on the left.  

\begin{figure}[ht]
	\includegraphics[width = 1\linewidth] {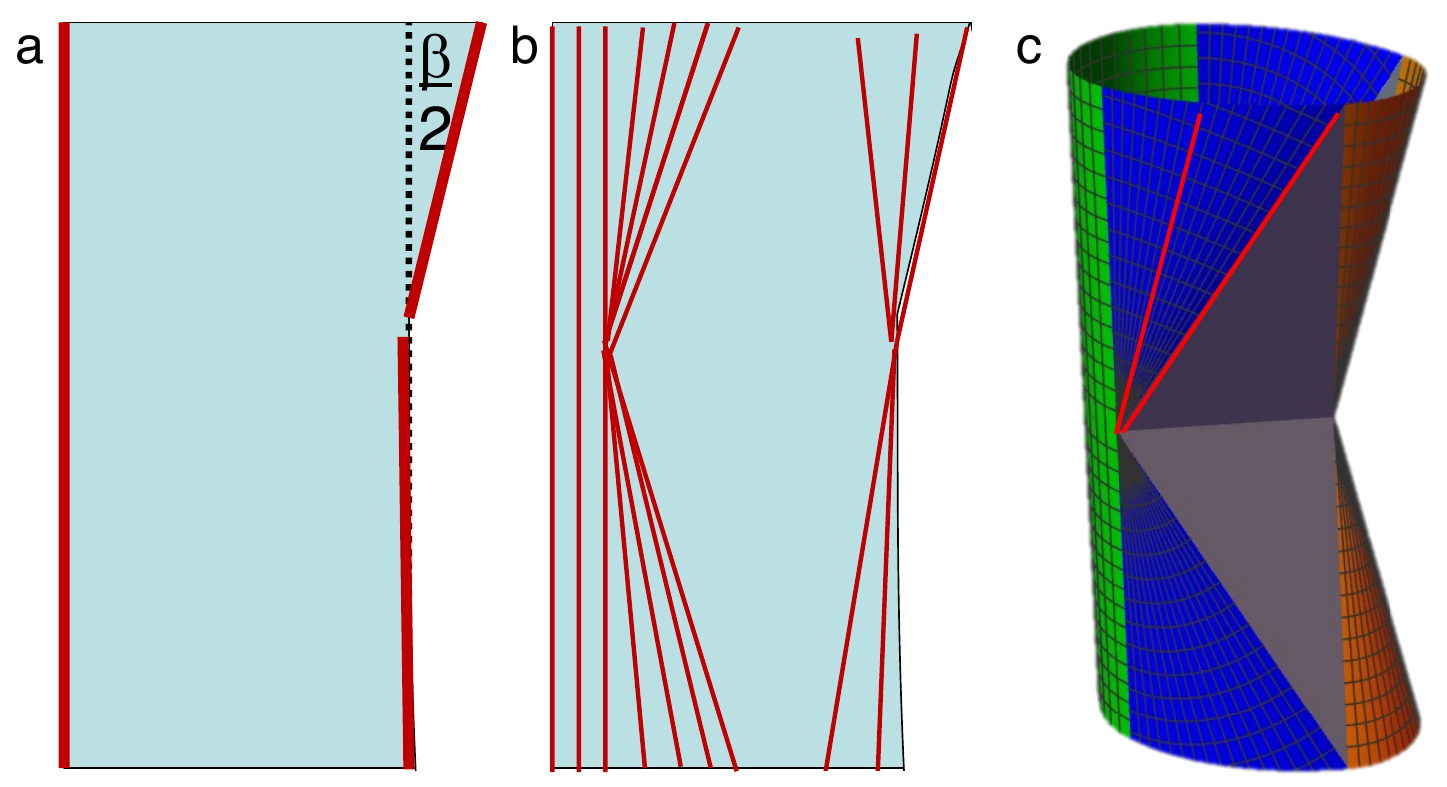}
    \caption {a)  Sketch of wedge-in-cylinder shape when flattened onto a plane.  b) Qualitative generator pattern corresponding to a smoothly curved developable surface.  c) Model construction showing an explicit curvature profile of the original flat sheet.  Here $\beta$-wedges were inserted symmetrically at top and bottom.  Green region on left has cylindrical curvature.   Gray triangular regions are flat facets. Two identical facets are hidden from view on the back side of the surface. Blue regions adjoining the cylinder are circular cone sectors.  Brown regions at right are circular cone sectors centered on the inserted wedges.
 }
    \label{fig:AnnaConstruction}
\end{figure}
The first step towards understanding this singularity is to determine the limiting developable structure when the thickness goes to zero.  It presumably consists of curved regions separated by flat facets, as in Fig. \ref{fig:boundaryridge}e.  From such a construction, one can understand what determines the observed angle between the facets and how it depends on the height of the cylinder.  How does the groove length and direction depend on height and the position of the wedge tip?  How moveable is the groove under boundary forces? Unlike the two-cone shapes of Fig.\ref{fig:boundaryridge}, determining this structure requires non-obvious calculation. Starting from the high-energy flat configuration of Fig. \ref{fig:AnnaConstruction}a, one may reduce the energy by spreading curvature to include more of the sheet, without violating isometry.  This means that new generator lines must emerge from the edges.  Those at the cylindrical left edge may expand to make vertical lines. Likewise, the right boundary may expand into a conical surface with converging generators, as sketched in Fig. \ref{fig:AnnaConstruction}b. The placement of these generators must be such that they describe bending of the original surface with no material added or removed.

Recently we have made a first step towards formulating this limiting developable shape, as sketched in Fig. \ref{fig:AnnaConstruction}c.   We made a symmetric variant with identical upper and lower wedges.  We designed circular cone sectors that smoothly bridged between the cylinder and the adjacent sheet.  Similarly, we allowed the region near the edges to bend into  circular cone sectors.  Matching the material between these cone segments requires a flat facet above and below, and these facets must be tilted relative to the cone axis whenever the cones' splay angle is nonzero.  Using this construction, we expect to find the facet tilt angle giving the smallest bending energy of the cones and cylinder. 

\section{Conclusion}
There's every reason to expect that the induced singular structures shown above can be understood on the same quantitative basis as the stretching ridge of Section \ref{sec:thinsheet}.  Our proposed scaling of ridge-induced boundary cones from Section \ref{sec:induced_bag} has yet to be validated.  No analogous predictions for most of the structures in Section \ref{sec:Menagerie} have even been formulated.  Still, we can hope for a systematic treatment of such sub-singularities.  This holds out the prospect of many new ways to control the shapes of sheets using external forcing.  

\begin{acknowledgments}
The unpublished work shown in Fig. \ref{fig:AnnaConstruction}c is the outgrowth of a multi-author research project \cite{Fleischer2023Emergent-remote}. We thank the participants in this project, especially Luka Pocivavsek and Efi Efrati, for pointing out the sub-singularity of Fig. \ref{fig:menagerie}d, and for many fruitful discussions.  We acknowledge fruitful conversations with Jos\'e Bico and Beno\^it Roman of ESPCI, Paris. This work received support from an internal grant from ``France and Chicago Collaborating in The Sciences". One of us (AM) was supported in part by the NSF MRSEC program under grant number DMR-2011854.
\end{acknowledgments}

% The \nocite command causes all entries in a bibliography to be printed out
% whether or not they are actually referenced in the text. This is appropriate
% for the sample file to show the different styles of references, but authors
% most likely will not want to use it.
\nocite{*}

\bibliography{GuyonPaper.bib}% Produces the bibliography via BibTeX.

%apsrev4-2.bst 2019-01-14 (MD) hand-edited version of apsrev4-1.bst
%Control: key (0)
%Control: author (8) initials jnrlst
%Control: editor formatted (1) identically to author
%Control: production of article title (0) allowed
%Control: page (0) single
%Control: year (1) truncated
%Control: production of eprint (0) enabled
\begin{thebibliography}{33}%
\makeatletter
\providecommand \@ifxundefined [1]{%
 \@ifx{#1\undefined}
}%
\providecommand \@ifnum [1]{%
 \ifnum #1\expandafter \@firstoftwo
 \else \expandafter \@secondoftwo
 \fi
}%
\providecommand \@ifx [1]{%
 \ifx #1\expandafter \@firstoftwo
 \else \expandafter \@secondoftwo
 \fi
}%
\providecommand \natexlab [1]{#1}%
\providecommand \enquote  [1]{``#1''}%
\providecommand \bibnamefont  [1]{#1}%
\providecommand \bibfnamefont [1]{#1}%
\providecommand \citenamefont [1]{#1}%
\providecommand \href@noop [0]{\@secondoftwo}%
\providecommand \href [0]{\begingroup \@sanitize@url \@href}%
\providecommand \@href[1]{\@@startlink{#1}\@@href}%
\providecommand \@@href[1]{\endgroup#1\@@endlink}%
\providecommand \@sanitize@url [0]{\catcode `\\12\catcode `\$12\catcode `\&12\catcode `\#12\catcode `\^12\catcode `\_12\catcode `\%12\relax}%
\providecommand \@@startlink[1]{}%
\providecommand \@@endlink[0]{}%
\providecommand \url  [0]{\begingroup\@sanitize@url \@url }%
\providecommand \@url [1]{\endgroup\@href {#1}{\urlprefix }}%
\providecommand \urlprefix  [0]{URL }%
\providecommand \Eprint [0]{\href }%
\providecommand \doibase [0]{https://doi.org/}%
\providecommand \selectlanguage [0]{\@gobble}%
\providecommand \bibinfo  [0]{\@secondoftwo}%
\providecommand \bibfield  [0]{\@secondoftwo}%
\providecommand \translation [1]{[#1]}%
\providecommand \BibitemOpen [0]{}%
\providecommand \bibitemStop [0]{}%
\providecommand \bibitemNoStop [0]{.\EOS\space}%
\providecommand \EOS [0]{\spacefactor3000\relax}%
\providecommand \BibitemShut  [1]{\csname bibitem#1\endcsname}%
\let\auto@bib@innerbib\@empty
%</preamble>
\bibitem [{\citenamefont {Pomeau}(1995)}]{pomeau1995papier}%
  \BibitemOpen
  \bibfield  {author} {\bibinfo {author} {\bibfnamefont {Y.}~\bibnamefont {Pomeau}},\ }\bibfield  {title} {\bibinfo {title} {Papier froiss{\'e}},\ }\href@noop {} {\bibfield  {journal} {\bibinfo  {journal} {Comptes rendus de l'Acad{\'e}mie des sciences. S{\'e}rie 1, Math{\'e}matique}\ }\textbf {\bibinfo {volume} {320}},\ \bibinfo {pages} {975} (\bibinfo {year} {1995})}\BibitemShut {NoStop}%
\bibitem [{\citenamefont {Lobkovsky}\ \emph {et~al.}(1995)\citenamefont {Lobkovsky}, \citenamefont {Gentges}, \citenamefont {Li}, \citenamefont {Morse},\ and\ \citenamefont {Witten}}]{Lobkovsky1995Scaling-Propert}%
  \BibitemOpen
  \bibfield  {author} {\bibinfo {author} {\bibfnamefont {A.}~\bibnamefont {Lobkovsky}}, \bibinfo {author} {\bibfnamefont {S.}~\bibnamefont {Gentges}}, \bibinfo {author} {\bibfnamefont {H.}~\bibnamefont {Li}}, \bibinfo {author} {\bibfnamefont {D.}~\bibnamefont {Morse}},\ and\ \bibinfo {author} {\bibfnamefont {T.~A.}\ \bibnamefont {Witten}},\ }\bibfield  {title} {\bibinfo {title} {Scaling properties of stretching ridges in a crumpled elastic sheet},\ }\href {https://doi.org/10.1126/science.270.5241.1482} {\bibfield  {journal} {\bibinfo  {journal} {Science}\ }\textbf {\bibinfo {volume} {270}},\ \bibinfo {pages} {1482} (\bibinfo {year} {1995})}\BibitemShut {NoStop}%
\bibitem [{\citenamefont {Amar}\ and\ \citenamefont {Pomeau}(1997)}]{Amar:1997fk}%
  \BibitemOpen
  \bibfield  {author} {\bibinfo {author} {\bibfnamefont {M.~B.}\ \bibnamefont {Amar}}\ and\ \bibinfo {author} {\bibfnamefont {Y.}~\bibnamefont {Pomeau}},\ }\bibfield  {title} {\bibinfo {title} {Crumpled paper},\ }\href {https://doi.org/10.1098/rspa.1997.0041} {\bibfield  {journal} {\bibinfo  {journal} {Proceedings of the Royal Society of London. Series A: Mathematical, Physical and Engineering Sciences}\ }\textbf {\bibinfo {volume} {453}},\ \bibinfo {pages} {729} (\bibinfo {year} {1997})}\BibitemShut {NoStop}%
\bibitem [{\citenamefont {Cerda}\ and\ \citenamefont {Mahadevan}(1998)}]{Cerda:1998cj}%
  \BibitemOpen
  \bibfield  {author} {\bibinfo {author} {\bibfnamefont {E.}~\bibnamefont {Cerda}}\ and\ \bibinfo {author} {\bibfnamefont {L.}~\bibnamefont {Mahadevan}},\ }\bibfield  {title} {\bibinfo {title} {Conical surfaces and crescent singularities in crumpled sheets},\ }\href {https://doi.org/10.1103/PhysRevLett.80.2358} {\bibfield  {journal} {\bibinfo  {journal} {Phys. Rev. Lett.}\ }\textbf {\bibinfo {volume} {80}},\ \bibinfo {pages} {2358} (\bibinfo {year} {1998})}\BibitemShut {NoStop}%
\bibitem [{\citenamefont {Py}\ \emph {et~al.}(2007)\citenamefont {Py}, \citenamefont {Reverdy}, \citenamefont {Doppler}, \citenamefont {Bico}, \citenamefont {Roman},\ and\ \citenamefont {Baroud}}]{Py:2007nr}%
  \BibitemOpen
  \bibfield  {author} {\bibinfo {author} {\bibfnamefont {C.}~\bibnamefont {Py}}, \bibinfo {author} {\bibfnamefont {P.}~\bibnamefont {Reverdy}}, \bibinfo {author} {\bibfnamefont {L.}~\bibnamefont {Doppler}}, \bibinfo {author} {\bibfnamefont {J.}~\bibnamefont {Bico}}, \bibinfo {author} {\bibfnamefont {B.}~\bibnamefont {Roman}},\ and\ \bibinfo {author} {\bibfnamefont {C.~N.}\ \bibnamefont {Baroud}},\ }\bibfield  {title} {\bibinfo {title} {Capillary origami: Spontaneous wrapping of a droplet with an elastic sheet},\ }\href {https://doi.org/10.1103/PhysRevLett.98.156103} {\bibfield  {journal} {\bibinfo  {journal} {Phys. Rev. Lett.}\ }\textbf {\bibinfo {volume} {98}},\ \bibinfo {pages} {156103} (\bibinfo {year} {2007})}\BibitemShut {NoStop}%
\bibitem [{\citenamefont {Roman}\ \emph {et~al.}(2003)\citenamefont {Roman}, \citenamefont {Reis}, \citenamefont {Audoly}, \citenamefont {De~Villiers}, \citenamefont {Vigui{\'e}},\ and\ \citenamefont {Vallet}}]{roman2003oscillatory}%
  \BibitemOpen
  \bibfield  {author} {\bibinfo {author} {\bibfnamefont {B.}~\bibnamefont {Roman}}, \bibinfo {author} {\bibfnamefont {P.~M.}\ \bibnamefont {Reis}}, \bibinfo {author} {\bibfnamefont {B.}~\bibnamefont {Audoly}}, \bibinfo {author} {\bibfnamefont {S.}~\bibnamefont {De~Villiers}}, \bibinfo {author} {\bibfnamefont {V.}~\bibnamefont {Vigui{\'e}}},\ and\ \bibinfo {author} {\bibfnamefont {D.}~\bibnamefont {Vallet}},\ }\bibfield  {title} {\bibinfo {title} {Oscillatory fracture paths in thin elastic sheets},\ }\href@noop {} {\bibfield  {journal} {\bibinfo  {journal} {Comptes Rendus Mecanique}\ }\textbf {\bibinfo {volume} {331}},\ \bibinfo {pages} {811} (\bibinfo {year} {2003})}\BibitemShut {NoStop}%
\bibitem [{\citenamefont {Si\'efert}\ \emph {et~al.}(2019)\citenamefont {Si\'efert}, \citenamefont {Reyssat}, \citenamefont {Bico},\ and\ \citenamefont {Roman}}]{Siefert:2019uo}%
  \BibitemOpen
  \bibfield  {author} {\bibinfo {author} {\bibfnamefont {E.}~\bibnamefont {Si\'efert}}, \bibinfo {author} {\bibfnamefont {E.}~\bibnamefont {Reyssat}}, \bibinfo {author} {\bibfnamefont {J.}~\bibnamefont {Bico}},\ and\ \bibinfo {author} {\bibfnamefont {B.}~\bibnamefont {Roman}},\ }\bibfield  {title} {\bibinfo {title} {Programming curvilinear paths of flat inflatables},\ }\href {https://doi.org/10.1073/pnas.1904544116} {\bibfield  {journal} {\bibinfo  {journal} {Proceedings of the National Academy of Sciences}\ }\textbf {\bibinfo {volume} {116}},\ \bibinfo {pages} {16692} (\bibinfo {year} {2019})}\BibitemShut {NoStop}%
\bibitem [{\citenamefont {Hutton}\ \emph {et~al.}(2024)\citenamefont {Hutton}, \citenamefont {Vitral}, \citenamefont {Hamm},\ and\ \citenamefont {Hanna}}]{Hutton2024Buckling-mediat}%
  \BibitemOpen
  \bibfield  {author} {\bibinfo {author} {\bibfnamefont {R.~S.}\ \bibnamefont {Hutton}}, \bibinfo {author} {\bibfnamefont {E.}~\bibnamefont {Vitral}}, \bibinfo {author} {\bibfnamefont {E.}~\bibnamefont {Hamm}},\ and\ \bibinfo {author} {\bibfnamefont {J.}~\bibnamefont {Hanna}},\ }\bibfield  {title} {\bibinfo {title} {Buckling mediated by mobile localized elastic excitations},\ }\bibfield  {journal} {\bibinfo  {journal} {PNAS Nexus}\ }\textbf {\bibinfo {volume} {3}},\ \href {https://doi.org/10.1093/pnasnexus/pgae083} {10.1093/pnasnexus/pgae083} (\bibinfo {year} {2024})\BibitemShut {NoStop}%
\bibitem [{\citenamefont {Cerda}\ \emph {et~al.}(1999)\citenamefont {Cerda}, \citenamefont {Chaieb}, \citenamefont {Melo},\ and\ \citenamefont {Mahadevan}}]{Cerda:1999fr}%
  \BibitemOpen
  \bibfield  {author} {\bibinfo {author} {\bibfnamefont {E.}~\bibnamefont {Cerda}}, \bibinfo {author} {\bibfnamefont {S.}~\bibnamefont {Chaieb}}, \bibinfo {author} {\bibfnamefont {F.}~\bibnamefont {Melo}},\ and\ \bibinfo {author} {\bibfnamefont {L.}~\bibnamefont {Mahadevan}},\ }\bibfield  {title} {\bibinfo {title} {Conical dislocations in crumpling},\ }\href {https://doi.org/10.1038/43395} {\bibfield  {journal} {\bibinfo  {journal} {Nature}\ }\textbf {\bibinfo {volume} {401}},\ \bibinfo {pages} {46 EP } (\bibinfo {year} {1999})}\BibitemShut {NoStop}%
\bibitem [{\citenamefont {Davidovitch}\ \emph {et~al.}(2011)\citenamefont {Davidovitch}, \citenamefont {Schroll}, \citenamefont {Vella}, \citenamefont {Adda-Bedia},\ and\ \citenamefont {Cerda}}]{Davidovitch:2011wq}%
  \BibitemOpen
  \bibfield  {author} {\bibinfo {author} {\bibfnamefont {B.}~\bibnamefont {Davidovitch}}, \bibinfo {author} {\bibfnamefont {R.~D.}\ \bibnamefont {Schroll}}, \bibinfo {author} {\bibfnamefont {D.}~\bibnamefont {Vella}}, \bibinfo {author} {\bibfnamefont {M.}~\bibnamefont {Adda-Bedia}},\ and\ \bibinfo {author} {\bibfnamefont {E.~A.}\ \bibnamefont {Cerda}},\ }\bibfield  {title} {\bibinfo {title} {Prototypical model for tensional wrinkling in thin sheets},\ }\href {https://doi.org/10.1073/pnas.1108553108} {\bibfield  {journal} {\bibinfo  {journal} {Proceedings of the National Academy of Sciences}\ }\textbf {\bibinfo {volume} {108}},\ \bibinfo {pages} {18227} (\bibinfo {year} {2011})},\ \Eprint {https://arxiv.org/abs/https://www.pnas.org/content/108/45/18227.full.pdf} {https://www.pnas.org/content/108/45/18227.full.pdf} \BibitemShut {NoStop}%
\bibitem [{\citenamefont {Bayart}\ \emph {et~al.}(2010)\citenamefont {Bayart}, \citenamefont {Boudaoud},\ and\ \citenamefont {Adda-Bedia}}]{bayart2010tearing}%
  \BibitemOpen
  \bibfield  {author} {\bibinfo {author} {\bibfnamefont {E.}~\bibnamefont {Bayart}}, \bibinfo {author} {\bibfnamefont {A.}~\bibnamefont {Boudaoud}},\ and\ \bibinfo {author} {\bibfnamefont {M.}~\bibnamefont {Adda-Bedia}},\ }\bibfield  {title} {\bibinfo {title} {On the tearing of thin sheets},\ }\href {https://doi.org/10.1016/j.engfracmech.2010.03.006} {\bibfield  {journal} {\bibinfo  {journal} {Engineering fracture mechanics}\ }\textbf {\bibinfo {volume} {77}},\ \bibinfo {pages} {1849} (\bibinfo {year} {2010})}\BibitemShut {NoStop}%
\bibitem [{\citenamefont {Aharoni}\ \emph {et~al.}(2017)\citenamefont {Aharoni}, \citenamefont {Todorova}, \citenamefont {Albarr{\'a}n}, \citenamefont {Goehring}, \citenamefont {Kamien},\ and\ \citenamefont {Katifori}}]{Aharoni2017The-smectic-ord}%
  \BibitemOpen
  \bibfield  {author} {\bibinfo {author} {\bibfnamefont {H.}~\bibnamefont {Aharoni}}, \bibinfo {author} {\bibfnamefont {D.~V.}\ \bibnamefont {Todorova}}, \bibinfo {author} {\bibfnamefont {O.}~\bibnamefont {Albarr{\'a}n}}, \bibinfo {author} {\bibfnamefont {L.}~\bibnamefont {Goehring}}, \bibinfo {author} {\bibfnamefont {R.~D.}\ \bibnamefont {Kamien}},\ and\ \bibinfo {author} {\bibfnamefont {E.}~\bibnamefont {Katifori}},\ }\bibfield  {title} {\bibinfo {title} {The smectic order of wrinkles},\ }\href {https://doi.org/10.1038/ncomms15809} {\bibfield  {journal} {\bibinfo  {journal} {Nature Communications}\ }\textbf {\bibinfo {volume} {8}},\ \bibinfo {pages} {15809} (\bibinfo {year} {2017})}\BibitemShut {NoStop}%
\bibitem [{\citenamefont {Hure}\ \emph {et~al.}(2011)\citenamefont {Hure}, \citenamefont {Roman},\ and\ \citenamefont {Bico}}]{Hure:2011fk}%
  \BibitemOpen
  \bibfield  {author} {\bibinfo {author} {\bibfnamefont {J.}~\bibnamefont {Hure}}, \bibinfo {author} {\bibfnamefont {B.}~\bibnamefont {Roman}},\ and\ \bibinfo {author} {\bibfnamefont {J.}~\bibnamefont {Bico}},\ }\bibfield  {title} {\bibinfo {title} {Wrapping an adhesive sphere with an elastic sheet},\ }\href {https://doi.org/10.1103/PhysRevLett.106.174301} {\bibfield  {journal} {\bibinfo  {journal} {Phys. Rev. Lett.}\ }\textbf {\bibinfo {volume} {106}},\ \bibinfo {pages} {174301} (\bibinfo {year} {2011})}\BibitemShut {NoStop}%
\bibitem [{\citenamefont {Pomeau}\ and\ \citenamefont {Rica}(1997)}]{pomeau1997plates}%
  \BibitemOpen
  \bibfield  {author} {\bibinfo {author} {\bibfnamefont {Y.}~\bibnamefont {Pomeau}}\ and\ \bibinfo {author} {\bibfnamefont {S.}~\bibnamefont {Rica}},\ }\bibfield  {title} {\bibinfo {title} {Plates under strong stress},\ }\href@noop {} {\bibfield  {journal} {\bibinfo  {journal} {Comptes Rendus de l'Academie des Sciences Series IIB Mechanics Physics Chemistry Astronomy}\ }\textbf {\bibinfo {volume} {4}},\ \bibinfo {pages} {181} (\bibinfo {year} {1997})}\BibitemShut {NoStop}%
\bibitem [{\citenamefont {Vandeparre}\ \emph {et~al.}(2011)\citenamefont {Vandeparre}, \citenamefont {Pi\~neirua}, \citenamefont {Brau}, \citenamefont {Roman}, \citenamefont {Bico}, \citenamefont {Gay}, \citenamefont {Bao}, \citenamefont {Lau}, \citenamefont {Reis},\ and\ \citenamefont {Damman}}]{Vandeparre:2011cs}%
  \BibitemOpen
  \bibfield  {author} {\bibinfo {author} {\bibfnamefont {H.}~\bibnamefont {Vandeparre}}, \bibinfo {author} {\bibfnamefont {M.}~\bibnamefont {Pi\~neirua}}, \bibinfo {author} {\bibfnamefont {F.}~\bibnamefont {Brau}}, \bibinfo {author} {\bibfnamefont {B.}~\bibnamefont {Roman}}, \bibinfo {author} {\bibfnamefont {J.}~\bibnamefont {Bico}}, \bibinfo {author} {\bibfnamefont {C.}~\bibnamefont {Gay}}, \bibinfo {author} {\bibfnamefont {W.}~\bibnamefont {Bao}}, \bibinfo {author} {\bibfnamefont {C.~N.}\ \bibnamefont {Lau}}, \bibinfo {author} {\bibfnamefont {P.~M.}\ \bibnamefont {Reis}},\ and\ \bibinfo {author} {\bibfnamefont {P.}~\bibnamefont {Damman}},\ }\bibfield  {title} {\bibinfo {title} {Wrinkling hierarchy in constrained thin sheets from suspended graphene to curtains},\ }\href {https://doi.org/10.1103/PhysRevLett.106.224301} {\bibfield  {journal} {\bibinfo  {journal} {Phys. Rev. Lett.}\ }\textbf {\bibinfo {volume} {106}},\ \bibinfo {pages} {224301} (\bibinfo {year} {2011})}\BibitemShut {NoStop}%
\bibitem [{\citenamefont {Duncan}\ and\ \citenamefont {Duncan}(1982)}]{Duncan:1982dw}%
  \BibitemOpen
  \bibfield  {author} {\bibinfo {author} {\bibfnamefont {J.~P.}\ \bibnamefont {Duncan}}\ and\ \bibinfo {author} {\bibfnamefont {J.~L.}\ \bibnamefont {Duncan}},\ }\bibfield  {title} {\bibinfo {title} {Folded developables},\ }\href {https://doi.org/10.1098/rspa.1982.0126} {\bibfield  {journal} {\bibinfo  {journal} {Proceedings of the Royal Society of London. A. Mathematical and Physical Sciences}\ }\textbf {\bibinfo {volume} {383}},\ \bibinfo {pages} {191} (\bibinfo {year} {1982})}\BibitemShut {NoStop}%
\bibitem [{\citenamefont {Meeussen}\ and\ \citenamefont {van Hecke}(2023)}]{Meeussen2023Multistable-she}%
  \BibitemOpen
  \bibfield  {author} {\bibinfo {author} {\bibfnamefont {A.~S.}\ \bibnamefont {Meeussen}}\ and\ \bibinfo {author} {\bibfnamefont {M.}~\bibnamefont {van Hecke}},\ }\bibfield  {title} {\bibinfo {title} {Multistable sheets with rewritable patterns for switchable shape-morphing},\ }\href {https://doi.org/10.1038/s41586-023-06353-5} {\bibfield  {journal} {\bibinfo  {journal} {Nature}\ }\textbf {\bibinfo {volume} {621}},\ \bibinfo {pages} {516} (\bibinfo {year} {2023})}\BibitemShut {NoStop}%
\bibitem [{\citenamefont {Tani}\ \emph {et~al.}(2024)\citenamefont {Tani}, \citenamefont {Hong}, \citenamefont {Tomizawa}, \citenamefont {Lepoivre}, \citenamefont {Bico},\ and\ \citenamefont {Roman}}]{tani2024curvy}%
  \BibitemOpen
  \bibfield  {author} {\bibinfo {author} {\bibfnamefont {M.}~\bibnamefont {Tani}}, \bibinfo {author} {\bibfnamefont {J.-W.}\ \bibnamefont {Hong}}, \bibinfo {author} {\bibfnamefont {T.}~\bibnamefont {Tomizawa}}, \bibinfo {author} {\bibfnamefont {{\'E}.}~\bibnamefont {Lepoivre}}, \bibinfo {author} {\bibfnamefont {J.}~\bibnamefont {Bico}},\ and\ \bibinfo {author} {\bibfnamefont {B.}~\bibnamefont {Roman}},\ }\bibfield  {title} {\bibinfo {title} {Curvy cuts: Programming axisymmetric kirigami shapes},\ }\href {https://doi.org/10.1016/j.eml.2024.102195} {\bibfield  {journal} {\bibinfo  {journal} {Extreme Mechanics Letters}\ }\textbf {\bibinfo {volume} {71}},\ \bibinfo {pages} {102195} (\bibinfo {year} {2024})}\BibitemShut {NoStop}%
\bibitem [{\citenamefont {Armon}\ \emph {et~al.}(2011)\citenamefont {Armon}, \citenamefont {Efrati}, \citenamefont {Kupferman},\ and\ \citenamefont {Sharon}}]{Armon2011Geometry-and-Me}%
  \BibitemOpen
  \bibfield  {author} {\bibinfo {author} {\bibfnamefont {S.}~\bibnamefont {Armon}}, \bibinfo {author} {\bibfnamefont {E.}~\bibnamefont {Efrati}}, \bibinfo {author} {\bibfnamefont {R.}~\bibnamefont {Kupferman}},\ and\ \bibinfo {author} {\bibfnamefont {E.}~\bibnamefont {Sharon}},\ }\bibfield  {title} {\bibinfo {title} {Geometry and mechanics in the opening of chiral seed pods},\ }\href {https://doi.org/10.1126/science.1203874} {\bibfield  {journal} {\bibinfo  {journal} {Science}\ }\textbf {\bibinfo {volume} {333}},\ \bibinfo {pages} {1726} (\bibinfo {year} {2011})},\ \Eprint {https://arxiv.org/abs/https://www.science.org/doi/pdf/10.1126/science.1203874} {https://www.science.org/doi/pdf/10.1126/science.1203874} \BibitemShut {NoStop}%
\bibitem [{\citenamefont {Vella}\ \emph {et~al.}(2015)\citenamefont {Vella}, \citenamefont {Huang}, \citenamefont {Menon}, \citenamefont {Russell},\ and\ \citenamefont {Davidovitch}}]{Vella:2015qy}%
  \BibitemOpen
  \bibfield  {author} {\bibinfo {author} {\bibfnamefont {D.}~\bibnamefont {Vella}}, \bibinfo {author} {\bibfnamefont {J.}~\bibnamefont {Huang}}, \bibinfo {author} {\bibfnamefont {N.}~\bibnamefont {Menon}}, \bibinfo {author} {\bibfnamefont {T.~P.}\ \bibnamefont {Russell}},\ and\ \bibinfo {author} {\bibfnamefont {B.}~\bibnamefont {Davidovitch}},\ }\bibfield  {title} {\bibinfo {title} {Indentation of ultrathin elastic films and the emergence of asymptotic isometry},\ }\bibfield  {journal} {\bibinfo  {journal} {Physical Review Letters}\ }\textbf {\bibinfo {volume} {114}},\ \href {https://doi.org/10.1103/physrevlett.114.014301} {10.1103/physrevlett.114.014301} (\bibinfo {year} {2015})\BibitemShut {NoStop}%
\bibitem [{\citenamefont {Sharon}\ \emph {et~al.}(2002)\citenamefont {Sharon}, \citenamefont {Roman}, \citenamefont {Marder}, \citenamefont {Shin},\ and\ \citenamefont {Swinney}}]{Sharon:2002kx}%
  \BibitemOpen
  \bibfield  {author} {\bibinfo {author} {\bibfnamefont {E.}~\bibnamefont {Sharon}}, \bibinfo {author} {\bibfnamefont {B.}~\bibnamefont {Roman}}, \bibinfo {author} {\bibfnamefont {M.}~\bibnamefont {Marder}}, \bibinfo {author} {\bibfnamefont {G.}~\bibnamefont {Shin}},\ and\ \bibinfo {author} {\bibfnamefont {H.}~\bibnamefont {Swinney}},\ }\bibfield  {title} {\bibinfo {title} {Mechanics: Buckling cascades in free sheets - wavy leaves may not depend only on their genes to make their edges crinkle.},\ }\href {https://doi.org/DOI 10.1038/419579a} {\bibfield  {journal} {\bibinfo  {journal} {Nature}\ }\textbf {\bibinfo {volume} {419}},\ \bibinfo {pages} {579} (\bibinfo {year} {2002})}\BibitemShut {NoStop}%
\bibitem [{\citenamefont {Box}\ \emph {et~al.}(2020)\citenamefont {Box}, \citenamefont {Kodio}, \citenamefont {O'Kiely}, \citenamefont {Cantelli}, \citenamefont {Goriely},\ and\ \citenamefont {Vella}}]{Box2020Dynamic-Bucklin}%
  \BibitemOpen
  \bibfield  {author} {\bibinfo {author} {\bibfnamefont {F.}~\bibnamefont {Box}}, \bibinfo {author} {\bibfnamefont {O.}~\bibnamefont {Kodio}}, \bibinfo {author} {\bibfnamefont {D.}~\bibnamefont {O'Kiely}}, \bibinfo {author} {\bibfnamefont {V.}~\bibnamefont {Cantelli}}, \bibinfo {author} {\bibfnamefont {A.}~\bibnamefont {Goriely}},\ and\ \bibinfo {author} {\bibfnamefont {D.}~\bibnamefont {Vella}},\ }\bibfield  {title} {\bibinfo {title} {Dynamic buckling of an elastic ring in a soap film},\ }\href {https://doi.org/10.1103/PhysRevLett.124.198003} {\bibfield  {journal} {\bibinfo  {journal} {Phys. Rev. Lett.}\ }\textbf {\bibinfo {volume} {124}},\ \bibinfo {pages} {198003} (\bibinfo {year} {2020})}\BibitemShut {NoStop}%
\bibitem [{\citenamefont {Pal}\ \emph {et~al.}(2024)\citenamefont {Pal}, \citenamefont {Pocivavsek},\ and\ \citenamefont {Witten}}]{Pal2024Faceted-wrinkli}%
  \BibitemOpen
  \bibfield  {author} {\bibinfo {author} {\bibfnamefont {A.~S.}\ \bibnamefont {Pal}}, \bibinfo {author} {\bibfnamefont {L.}~\bibnamefont {Pocivavsek}},\ and\ \bibinfo {author} {\bibfnamefont {T.~A.}\ \bibnamefont {Witten}},\ }\bibfield  {title} {\bibinfo {title} {Faceted wrinkling by contracting a curved boundary},\ }\bibfield  {journal} {\bibinfo  {journal} {Soft Matter}\ }\href {https://doi.org/10.1039/d3sm01347b} {10.1039/d3sm01347b} (\bibinfo {year} {2024})\BibitemShut {NoStop}%
\bibitem [{\citenamefont {He}\ \emph {et~al.}(2023)\citenamefont {He}, \citenamefont {D{\'e}mery},\ and\ \citenamefont {Paulsen}}]{He:2023rm}%
  \BibitemOpen
  \bibfield  {author} {\bibinfo {author} {\bibfnamefont {M.}~\bibnamefont {He}}, \bibinfo {author} {\bibfnamefont {V.}~\bibnamefont {D{\'e}mery}},\ and\ \bibinfo {author} {\bibfnamefont {J.~D.}\ \bibnamefont {Paulsen}},\ }\bibfield  {title} {\bibinfo {title} {Cross-sections of doubly curved sheets as confined elastica},\ }\href {https://doi.org/10.1073/pnas.2216786120} {\bibfield  {journal} {\bibinfo  {journal} {Proceedings of the National Academy of Sciences}\ }\textbf {\bibinfo {volume} {120}},\ \bibinfo {pages} {e2216786120} (\bibinfo {year} {2023})},\ \Eprint {https://arxiv.org/abs/https://www.pnas.org/doi/pdf/10.1073/pnas.2216786120} {https://www.pnas.org/doi/pdf/10.1073/pnas.2216786120} \BibitemShut {NoStop}%
\bibitem [{\citenamefont {Tobasco}\ \emph {et~al.}(2022)\citenamefont {Tobasco}, \citenamefont {Timounay}, \citenamefont {Todorova}, \citenamefont {Leggat}, \citenamefont {Paulsen},\ and\ \citenamefont {Katifori}}]{Tobasco:2022xe}%
  \BibitemOpen
  \bibfield  {author} {\bibinfo {author} {\bibfnamefont {I.}~\bibnamefont {Tobasco}}, \bibinfo {author} {\bibfnamefont {Y.}~\bibnamefont {Timounay}}, \bibinfo {author} {\bibfnamefont {D.}~\bibnamefont {Todorova}}, \bibinfo {author} {\bibfnamefont {G.~C.}\ \bibnamefont {Leggat}}, \bibinfo {author} {\bibfnamefont {J.~D.}\ \bibnamefont {Paulsen}},\ and\ \bibinfo {author} {\bibfnamefont {E.}~\bibnamefont {Katifori}},\ }\bibfield  {title} {\bibinfo {title} {Exact solutions for the wrinkle patterns of confined elastic shells},\ }\href {https://doi.org/10.1038/s41567-022-01672-2} {\bibfield  {journal} {\bibinfo  {journal} {Nature Physics}\ }\textbf {\bibinfo {volume} {18}},\ \bibinfo {pages} {1099} (\bibinfo {year} {2022})}\BibitemShut {NoStop}%
\bibitem [{\citenamefont {Sharon}\ \emph {et~al.}(2007)\citenamefont {Sharon}, \citenamefont {Roman},\ and\ \citenamefont {Swinney}}]{Sharon:2007le}%
  \BibitemOpen
  \bibfield  {author} {\bibinfo {author} {\bibfnamefont {E.}~\bibnamefont {Sharon}}, \bibinfo {author} {\bibfnamefont {B.}~\bibnamefont {Roman}},\ and\ \bibinfo {author} {\bibfnamefont {H.~L.}\ \bibnamefont {Swinney}},\ }\bibfield  {title} {\bibinfo {title} {Geometrically driven wrinkling observed in free plastic sheets and leaves},\ }\href {https://doi.org/10.1103/PhysRevE.75.046211} {\bibfield  {journal} {\bibinfo  {journal} {Phys. Rev. E}\ }\textbf {\bibinfo {volume} {75}},\ \bibinfo {pages} {046211} (\bibinfo {year} {2007})}\BibitemShut {NoStop}%
\bibitem [{\citenamefont {Kim}\ \emph {et~al.}(2012)\citenamefont {Kim}, \citenamefont {Hanna}, \citenamefont {Byun}, \citenamefont {Santangelo},\ and\ \citenamefont {Hayward}}]{Kim2012Designing-Respo}%
  \BibitemOpen
  \bibfield  {author} {\bibinfo {author} {\bibfnamefont {J.}~\bibnamefont {Kim}}, \bibinfo {author} {\bibfnamefont {J.~A.}\ \bibnamefont {Hanna}}, \bibinfo {author} {\bibfnamefont {M.}~\bibnamefont {Byun}}, \bibinfo {author} {\bibfnamefont {C.~D.}\ \bibnamefont {Santangelo}},\ and\ \bibinfo {author} {\bibfnamefont {R.~C.}\ \bibnamefont {Hayward}},\ }\bibfield  {title} {\bibinfo {title} {Designing responsive buckled surfaces by halftone gel lithography},\ }\href {https://doi.org/10.1126/science.1215309} {\bibfield  {journal} {\bibinfo  {journal} {Science}\ }\textbf {\bibinfo {volume} {335}},\ \bibinfo {pages} {1201} (\bibinfo {year} {2012})},\ \Eprint {https://arxiv.org/abs/https://www.science.org/doi/pdf/10.1126/science.1215309} {https://www.science.org/doi/pdf/10.1126/science.1215309} \BibitemShut {NoStop}%
\bibitem [{\citenamefont {Witten}(2007)}]{Witten:2007wq}%
  \BibitemOpen
  \bibfield  {author} {\bibinfo {author} {\bibfnamefont {T.~A.}\ \bibnamefont {Witten}},\ }\bibfield  {title} {\bibinfo {title} {Stress focusing in elastic sheets},\ }\href {https://doi.org/10.1103/RevModPhys.79.643} {\bibfield  {journal} {\bibinfo  {journal} {Rev. Mod. Phys.}\ }\textbf {\bibinfo {volume} {79}},\ \bibinfo {pages} {643} (\bibinfo {year} {2007})}\BibitemShut {NoStop}%
\bibitem [{\citenamefont {Landau}\ and\ \citenamefont {Lifshitz}(1986)}]{Landau:1986sf}%
  \BibitemOpen
  \bibfield  {author} {\bibinfo {author} {\bibfnamefont {L.~D.}\ \bibnamefont {Landau}}\ and\ \bibinfo {author} {\bibfnamefont {E.~M.}\ \bibnamefont {Lifshitz}},\ }\href@noop {} {\emph {\bibinfo {title} {Theory of elasticity: volume 7}}},\ \bibinfo {edition} {3rd}\ ed.,\ edited by\ \bibinfo {editor} {\bibfnamefont {A.~M.}\ \bibnamefont {Kosevich}}\ and\ \bibinfo {editor} {\bibfnamefont {L.~P.}\ \bibnamefont {Pitaevskii}},\ Vol.~\bibinfo {volume} {7}\ (\bibinfo  {publisher} {Elsevier},\ \bibinfo {address} {Amsterdam},\ \bibinfo {year} {1986})\BibitemShut {NoStop}%
\bibitem [{\citenamefont {Witten}(2009)}]{Witten:2009kq}%
  \BibitemOpen
  \bibfield  {author} {\bibinfo {author} {\bibfnamefont {T.~A.}\ \bibnamefont {Witten}},\ }\bibfield  {title} {\bibinfo {title} {Spontaneous free-boundary structure in crumpled membranes},\ }\href {https://doi.org/10.1021/jp807548s} {\bibfield  {journal} {\bibinfo  {journal} {The Journal of Physical Chemistry B}\ }\textbf {\bibinfo {volume} {113}},\ \bibinfo {pages} {3738} (\bibinfo {year} {2009})}\BibitemShut {NoStop}%
\bibitem [{\citenamefont {Lobkovsky}(1996{\natexlab{a}})}]{Lobkovsky1996Structure-of-Cr}%
  \BibitemOpen
  \bibfield  {author} {\bibinfo {author} {\bibfnamefont {A.~E.}\ \bibnamefont {Lobkovsky}},\ }\emph {\bibinfo {title} {Structure of Crumpled Thin Elastic Membranes}},\ \href {https://jfi.uchicago.edu/~tten/Crumpling/Lobkovsky.thesis.pdf} {\bibinfo {type} {Phd dissertation}},\ \bibinfo  {school} {University of Chicago}, \bibinfo {address} {Chicago, IL} (\bibinfo {year} {1996}{\natexlab{a}})\BibitemShut {NoStop}%
\bibitem [{\citenamefont {Lobkovsky}(1996{\natexlab{b}})}]{Lobkovsky1996Boundary}%
  \BibitemOpen
  \bibfield  {author} {\bibinfo {author} {\bibfnamefont {A.~E.}\ \bibnamefont {Lobkovsky}},\ }\bibfield  {title} {\bibinfo {title} {Boundary layer analysis of the ridge singularity in a thin plate},\ }\href {https://doi.org/10.1103/PhysRevE.53.3750} {\bibfield  {journal} {\bibinfo  {journal} {Phys. Rev. E}\ }\textbf {\bibinfo {volume} {53}},\ \bibinfo {pages} {3750} (\bibinfo {year} {1996}{\natexlab{b}})}\BibitemShut {NoStop}%
\bibitem [{\citenamefont {Fleischer}\ \emph {et~al.}(2023)\citenamefont {Fleischer}, \citenamefont {Nguyen}, \citenamefont {Pal}, \citenamefont {Kim}, \citenamefont {Khabaz}, \citenamefont {Tsamis}, \citenamefont {Efrati}, \citenamefont {Witten}, \citenamefont {Alverdy},\ and\ \citenamefont {Pocivavsek}}]{Fleischer2023Emergent-remote}%
  \BibitemOpen
  \bibfield  {author} {\bibinfo {author} {\bibfnamefont {B.~P.}\ \bibnamefont {Fleischer}}, \bibinfo {author} {\bibfnamefont {N.}~\bibnamefont {Nguyen}}, \bibinfo {author} {\bibfnamefont {A.}~\bibnamefont {Pal}}, \bibinfo {author} {\bibfnamefont {J.}~\bibnamefont {Kim}}, \bibinfo {author} {\bibfnamefont {K.}~\bibnamefont {Khabaz}}, \bibinfo {author} {\bibfnamefont {A.}~\bibnamefont {Tsamis}}, \bibinfo {author} {\bibfnamefont {E.}~\bibnamefont {Efrati}}, \bibinfo {author} {\bibfnamefont {T.}~\bibnamefont {Witten}}, \bibinfo {author} {\bibfnamefont {J.~C.}\ \bibnamefont {Alverdy}},\ and\ \bibinfo {author} {\bibfnamefont {L.}~\bibnamefont {Pocivavsek}},\ }\bibfield  {title} {\bibinfo {title} {Emergent remote stress-focusing drive submucosal collagen fiber remodeling in intestinal anastomotic surgery},\ }\bibfield  {journal} {\bibinfo  {journal} {bioRxiv}\ }\href {https://doi.org/10.1101/2022.10.31.514540} {10.1101/2022.10.31.514540} (\bibinfo {year} {2023})\BibitemShut {NoStop}%
\end{thebibliography}%

\end{document}